\documentclass[twocolumn]{aastex6}

\usepackage{morefloats,url}
\usepackage{graphicx}
\usepackage{amssymb}
\usepackage{ mathrsfs }
\usepackage{amssymb}
\usepackage{amsmath}
\usepackage{cases}

\def\apjl{ApJ}%
\def\apjs{ApJS}%
\def\ao{Appl.~Opt.}%
\def\aap{A\&A}%
\usepackage{tabularx}

\slugcomment{Draft Version}

\shorttitle{Discovery of an eccentric MSP with a LMWD companion}
\shortauthors{J. Antoniadis {\it et~al.}}

\def\psr{PSR\,J2234+0611~}

\begin{document}

\title{An eccentric binary millisecond pulsar with a helium white dwarf companion in the Galactic field}

\author{John Antoniadis}
\affil{Dunlap Institute for Astronomy \& Astrophysics, University of Toronto, 50 St. George Street
Toronto, Ontario Canada M5S 3H4}
\email{antoniadis@dunlap.utoronto.ca}

\and

\author{David L. Kaplan}
\affil{Department of Physics, University of Wisconsin-Milwaukee, 1900 East Kenwood Boulevard, Milwaukee, WI 53211}

\and

\author{Kevin Stovall}
\affil{Department of Physics and Astronomy, University of New Mexico, Albuquerque, NM, 87131, USA}

\and

\author{Paulo C. C. Freire}
\affil{Max-Planck-Institut f\"{u}r Radioastronomie, Auf dem H\"{u}gel 69, 53121 Bonn, Germany}

\and

\author{Julia S. Deneva}
\affil{National Research Council, resident at the Naval Research Laboratory,4555 Overlook Ave SW, Washington, DC, 20375}

\and

\author{Detlev Koester}
\affil{Institut f\"{u}r Theoretische Physik und Astrophysik, Christian-Albrechts-Univerist\"{a}t Kiel, 24098 Kiel, Germany}

\and

\author{Fredrick Jenet}
\affil{Center for Advanced Radio Astronomy, Department of Physics and Astronomy, University of Texas at Brownsville, Brownsville, \\TX 78520}

\and 

\author{Jose G. Martinez}
\affil{Max-Planck-Institut f\"{u}r Radioastronomie, Auf dem H\"{u}gel 69, 53121 Bonn, Germany}

\begin{abstract} 
Low-mass  white dwarfs (LMWDs)  are believed to be exclusive products of binary evolution, as 
the Universe is not old enough to produce them from single stars.  Because of the strong tidal 
forces operating  during the binary interaction phase,  the remnant  systems observed today are 
expected to have negligible eccentricities.  Here, we report on the first unambiguous 
identification of a LMWD in an eccentric ($e=0.13$) orbit around the millisecond pulsar PSR\,J2234+0511, which directly  contradicts this picture.  We use our spectra and radio-timing 
solution (derived elsewhere) to infer the WD temperature ($T_{\rm eff} = 8600\pm190$\,K), and 
peculiar systemic velocity relative to the local standard of rest ($\simeq31$\,km\,s$^{-1}$). We 
also place model-independent constraints on the WD radius ($R_{\rm WD} = 0.024^{+0.004}_{-0.002}$\,R$_{\odot}$) and surface gravity ($\log g = 7.11^{+0.08}_{-0.16}$\,dex).  The WD 
and kinematic properties are consistent with the expectations for low-mass {X}-ray binary 
evolution and disfavour a dynamic three-body formation channel.   In the case of the high 
eccentricity being the result of a spontaneous phase transition, we infer a  mass of $\sim 1.60$\,M$_{\odot}$ for the pulsar progenitor, which is too low for the quark-nova mechanism proposed 
by Jiang et al. (2015), and too high for the scenario of Freire \& Tauris (2014), in which a WD 
collapses into a neutron star via a rotationally-delayed accretion-induced collapse. We find that 
eccentricity pumping via interaction with  a circumbinary disk is consistent with our inferred 
parameters.  Finally, we report tentative evidence for pulsations which, if confirmed, would 
transform the star into an unprecedented laboratory for WD physics.

\end{abstract}

\section{Introduction} 
Millisecond radio pulsars (MSPs) are a distinct population of neutron stars (NSs) with 
fast spin periods and magnetic fields several orders of magnitude weaker than those of ``normal'' pulsars. 
These properties are thought to reflect  a long mass transfer episode from a stellar companion. 
In the standard ``recycling'' scenario, the donor's mass accumulated 
on the NS increases its spin frequency and, through a process not yet 
well understood, buries the magnetic field \citep[see][]{tv86,bvh91}. 
Depending on the orbital period and initial donor mass, 
the remnant companions are usually low-mass white dwarfs (LMWDs; with $0.16 - 0.4$\,M$_{\odot}$), or semi-degenerate stars with masses in the range $0.001 - 0.4$\,M$_{\odot}$ \citep[e.g.][]{asr+09,tau11}. 
As a result of efficient tidal dissipation during the mass-transfer episode, 
the post-interaction orbits are expected to be extremely circular ($e \sim 10^{-7} - 10^{-3}$), which is indeed the case for most MSPs in the Galactic field \citep{phi92}. 

Surprisingly, recent surveys with the Arecibo,  Effelsberg and Parkes radio 
telescopes, led to the discovery of five peculiar binary MSPs with mass functions indicative of low-mass companions,  
 but large eccentricities of $e \simeq 0.025 - 0.13$ (henceforth eMSPs).  
The only previously known eccentric MSP in the Galactic field, 
 PSR\,J1903+0327 \citep{crl+08,fbw+11} has a  substantially 
higher eccentricity ($e\simeq0.44$) and a 1\,M$_{\odot}$ main-sequence companion. 
As the latter could not have been responsible for recycling the pulsar, 
the binary is most likely the remnant of a triple, 
where the least massive star, responsible for recycling the MSP, was ejected as a result of unstable orbital evolution \citep{fbw+11,pvv11,bfhz11,pk12,pcp12}.

The recently identified eMSPs are qualitatively different from PSR\,J1903+0327. 
The binary companions have very low masses, suggesting that they most likely descend from the  donors that spun-up the pulsars. 
In addition, their orbits resemble each other in many ways, making an origin in a triple system unlikely, because the chaotic disruption of the original systems would also result in a broad range of orbital properties for the observed remnant binaries.

An  attractive alternative scenario is that eMSPs are the 
direct descendants of low-mass {X}-ray binaries (LMXBs). 
This  hypothesis places strong constraints both on 
the expected observational properties and on the formation mechanism. 
For LMXBs with sufficiently large initial separations, 
the recycling episode starts when the companion 
evolves off the main sequence. In red giants, the mass of the helium core is proportional to the   
stellar radius, which in turn  is regulated by the orbital separation. Therefore, the final companions are expected to be  LMWDs with  masses proportional to the orbital period \citep{ts99a}. 
In addition, because tidal dissipation acts on sub-thermal timescales, the eccentricities must have increased either {\it after} or during the very last phase of the long-term recycling episode. 

A possible explanation for the observed eccentricities is a spontaneous phase transition of either a super-Chandrasekhar WD collapsing into a NS \citep{ft14}, or of a NS imploding into a strange-quark star \citep{jldd15}. In either case, the transformation is mass-critical and therefore the pulsar masses should be similar. An alternative mechanism could be the interaction of the post-LMXB system with a circumbinary disk \citep{ant14}, in which case  the 
pulsar masses need not be similar. A comprehensive comparison of the aforementioned scenarios and their predictions   is given in Table\,1 and discussed further in Section\,4. 

In this work we present  optical observations of PSR\,J2234+0611 \citep{dsm+13}, a nearby eMSP with a 32-day orbital period and an eccentricity of 0.13.   Our spectroscopy unambiguously  confirms that the companion is a LMWD, making PSR\,J2234+0611 the first such system identified in the Galactic field. The paper is structured as follows: in Section\,2 we describe our dataset and analysis. In Section\,3 we present constraints on the WD and NS mass and finally, in Section\,4 we explore the ramifications for the proposed formation theories. We conclude in Section\,5. 

\begin{table*}[!ht]
\caption{Comparison of theories for the formation of eccentric MSPs}
\centering 
\begin{tabularx}{\textwidth}{lXXXXXX}
\hline
\textbf{Theory} &\textbf{ Orbital Periods} &\textbf{ Companion } & \textbf{Companion Mass}& \textbf{Pulsar Mass }& \textbf{Notes }& \textbf{Reference} \\

\hline \hline
Triple System &unbound & MS or WD & unbound & unbound & & \\[10px]
Delayed AIC & $10-60$\,days & LMWD & 0.25-0.35\,M$_{\odot}$ & $1.35$\,M$_{\odot}$ & The implosion is symmetric, leading to small systemic velocities & \cite{ft14} \\[10px]

Quark phase transition & unbound & LMWD & unbound & $\sim 1.8$\,M$_{\odot}$ & & \cite{jldd15} \\[10px]
CB Disk & $15 -30$\,days& LMWD & 0.22-0.35\,M$_{\odot}$ & unbound & Small number of circular binaries within this orbital period range &  \cite{ant14}\\

\hline 

\end{tabularx}
\end{table*}

\section{Observations}
\subsection{Radio Timing}
The radio timing of the pulsar is presented in a companion paper by Stovall et al. (in prep; Paper\,II). 
Here, we summarise some of the key results for completeness.

The timing baseline, which now spans over two years, allows for a precise determination of the proper motion, [$\mu_{\alpha},\mu_{\delta}] = [25.390(27),9.48(0.06)$]\,mas\,yr$^{-1}$, and the parallax, $\varpi = 0.742(28)$\,mas, placing the system at a distance of $1.35(5)$\,kpc. 
In addition, the combined constraints on  relativistic periastron precession and  Shapiro delay, yield a mass of $M_{\rm c} = 0.275(8)$\,M$_{\odot}$ for the companion and $M_{\rm PSR}=1.39(1)$\,M$_{\odot}$ for the pulsar. In Sections 3 \& 4, we use these measurements to constrain  the size of the companion, compute the orbit of the system in the Galaxy, and  explore the implications for the proposed formation scenarios.

\subsection{Optical Observations}
The optical counterpart to  PSR\,J2234+0611 was first identified in the 
SDSS archive \citep{sdss} as a faint star ($g = 22.17$) coincident with timing position of the pulsar. 
Given the small stellar density of the field, the probability for a chance alignment is low ($< 0.001\%$). Based on the measured SDSS $ugriz$ magnitudes, the parallax distance, and 
a collection of WD cooling models  \citep[e.g.][]{sar+01,afw+13}, we find that the star is consistent with being a LMWD with a temperature  of $T_{\rm eff}\simeq 8500$\,K.

 We followed up PSR\,J2234+0611  spectroscopically using the FORS2 instrument \citep{fors}  
of the Very Large Telescope (VLT) in Chile. All observations were carried out in service mode between July and September 2014 using the MIT red-sensitive camera  which delivers a resolution of 0.25\,arcsec per binned-by-two pixel along the spatial direction. 
 We used the 1200 lines\,mm$^{-1}$ grism (GRIS$\_1200$B+97) which covers the spectral range between 378 and 510\,nm with a resolution of 0.072\,nm\,pix$^{-1}$. 
 A total of 26, 1420\,sec spectra were collected through a 1-arcsec slit, which was rotated by 
 $31.3$ degrees (North to East) in respect to the parallactic angle to include a bright reference star south-west of the target. In addition, we recorded two 1300\,s exposures using a 5-arcsec slit for calibration purposes. The atmospheric conditions were generally variable, but some exposures were taken during clear conditions. The seeing ranged from 0.6 to 1.5 arcsec. 

The data were reduced using astropy \citep{astropy}\footnote{http://www.astropy.org} and custom python routines. First, we removed the bias level using the median estimate from the overscan region.  Then, we corrected for small-scale sensitivity variations using  lamp exposures, normalized both along the spatial and the dispersion directions. The sky level was estimated by using a second-degree polynomial fitted to 100-arcsec  regions on either side of the target and the reference, but excluding  5-arcsec areas around them. Finally, we extracted the spectra using an optimal-weighting method similar to that of \cite{h86}. 

The dispersion solution was derived using a second-degree polynomial fit to the {CuAr} lamp spectra,  recorded at the end of each run. The solutions have  root-mean-square (rms) residuals  ranging from $\sim 0.02$ to 0.04\,nm for 18 to 20 lines. 
For flux calibration, we first corrected for wavelength-dependent slit losses using the normalized wide-to-narrow slit flux ratio of the reference star and then applied the instrument response derived using an exposure of a standard star on a photometric night.  

\section{Results}

\subsection{Radial velocities and constraints on the mass ratio} 
Radial velocities were measured using the method described in \cite{avk+12}. 
First, we identified the reference star as being a K3V star and used a medium-resolution PHOENIX spectrum as a template to measure its velocity \citep{phoenix}. For PSR\,J2234+0611, we first fitted  a grid of DA model atmospheres \citep{spectra} to a single spectrum, used the best-fit model as a template to shift and average the spectra, and then refined the template by fitting the average spectrum. 

\begin{figure}
\begin{center}
\includegraphics[width=0.45\textwidth]{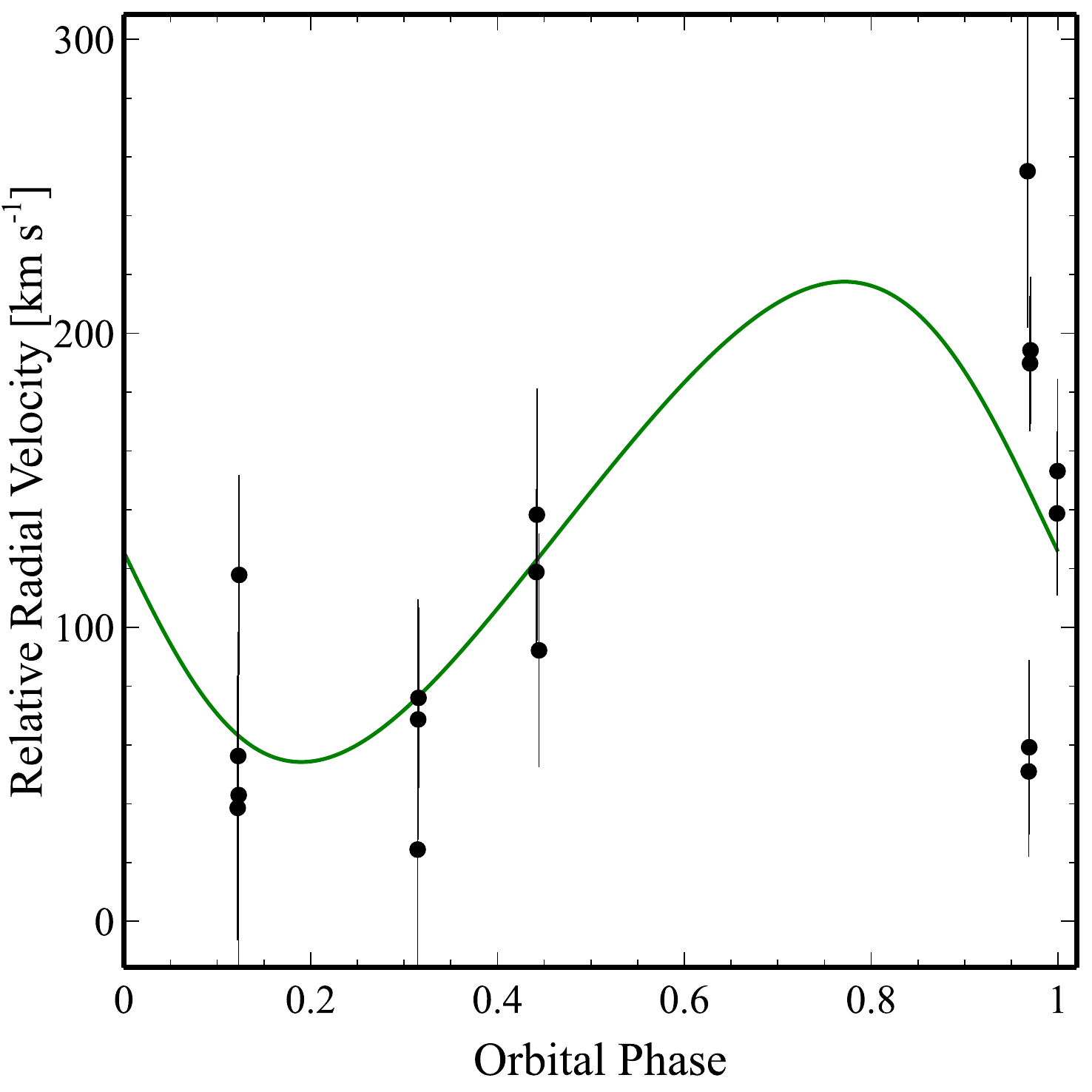}
\caption{Radial velocities of the companion to PSR\,J2234+0611 as a function of the orbital phase. The solid line shows the inferred best-fit orbit.}
\end{center}
\end{figure}

Figure\,1 shows \deleted{our radial velocity measurements as a function of  orbital phase.} \added{the radial velocities of the WD relative to that of the reference star, as a function of orbital phase.} Unfortunately, the uncertainties of individual measurements ($> 23$\,km\,s$^{-1}$) are  comparable to the expected amplitude of the orbital velocity of $K_{\rm c} = 1.39/0.275  2\pi x/P_{\rm b} = 47.99$\,km\,s$^{-1}$, thereby not allowing for a precise independent  determination of the mass ratio.
After excluding 8 measurements with uncertainties larger than $60$\,km\,s$^{-1}$ and an extreme outlier with $\delta v = 442\pm22$\,km\,s$^{-1}$, a fit to the orbit using the timing ephemeris of the pulsar ---with 
$T_{0} = 56794.0931864$\,MJD, $P_{\rm b} = 32.00140$\,d, $e = 0.129274$  and $\omega = 277.16764$\,degrees--- yields an  orbital velocity semi-amplitude of 
$K_{\rm c} = 81\pm23$\,km\,s$^{-1}$ and a systemic velocity of $\Delta \gamma$= +137.2\,km\,s$^{-1}$ relative to the reference star, with $\chi_{\rm red} = 1.9$ for 15 degrees of freedom. Combined with the orbital parameters of the pulsar constrained with timing, the inferred  mass ratio is $q \equiv M_{\rm PSR}/M_{\rm c} = 8.5\pm2.4$. 

The radial velocity of the reference star varied as much as 44\,km\,s$^{-1}$ among the 17 observations, which is considerably larger than the formal uncertainties. However, \added{for the latter}, we find no evidence for binarity. A visual inspection of the through-slit pointing frames taken before the science exposures  suggests that the most likely cause is differential diffraction due to minor displacements of the star inside the slit. 
The inferred mean radial velocity is $\gamma_{\rm ref} = -93\pm 14$\,km\,s$^{-1}$, which differs by $25$\,km\,s$^{-1}$ compared  to the velocity found from the wide-slit spectra, $\gamma_{\rm ref, wide} = -117\pm 4$\,km\,s$^{-1}$. Conservatively adopting $\gamma_{\rm ref} = -117\pm 25$\,km\,s$^{-1}$, yields an absolute systemic velocity of $\gamma = 20\pm34$\,km\,s$^{-1}$ for PSR\,J2234+0611, \added{relative to the Solar system barycentre}. 

\subsection{Atmospheric parameters}
The zero-velocity average of the 17 best spectra mentioned above  is shown in Figure\,2. The spectrum is that of a typical pure-hydrogen WD, thereby confirming our initial photometric classification.  

We fitted the spectrum using a grid of DA model atmospheres covering the range $6000-10000$\,K with a stepsize of 250\,K in $T_{\rm eff}$, and $6.0-8.0$\,dex with a stepsize of $0.25$\,dex in $\log g$ \citep{spectra}. For the fit, we convolved each model using a Gaussian kernel with a dispersion equal to that of the average seeing truncated at the slit width. We also allowed for the normalization to vary with wavelength using a third-degree polynomial.  
The fit yields an effective temperature of $T_{\rm eff} = 8749\pm120$\,K and a surface gravity of $\log g = 7.25\pm 0.15$\,dex. To get an estimate for the systematics we varied the degree of the normalization polynomial, the parameters of the convolution kernel and the spectral range used for the fit. These fits were overall consistent with each other, with a scatter slightly larger than the 1$\sigma$ formal uncertainties. 
To compensate for that we adopt $T_{\rm eff} = 8750\pm200$\,K and $\log g = 7.25\pm 0.20$\,dex for all our calculations below. 

\begin{figure}
\begin{center}
\includegraphics[width=0.45\textwidth]{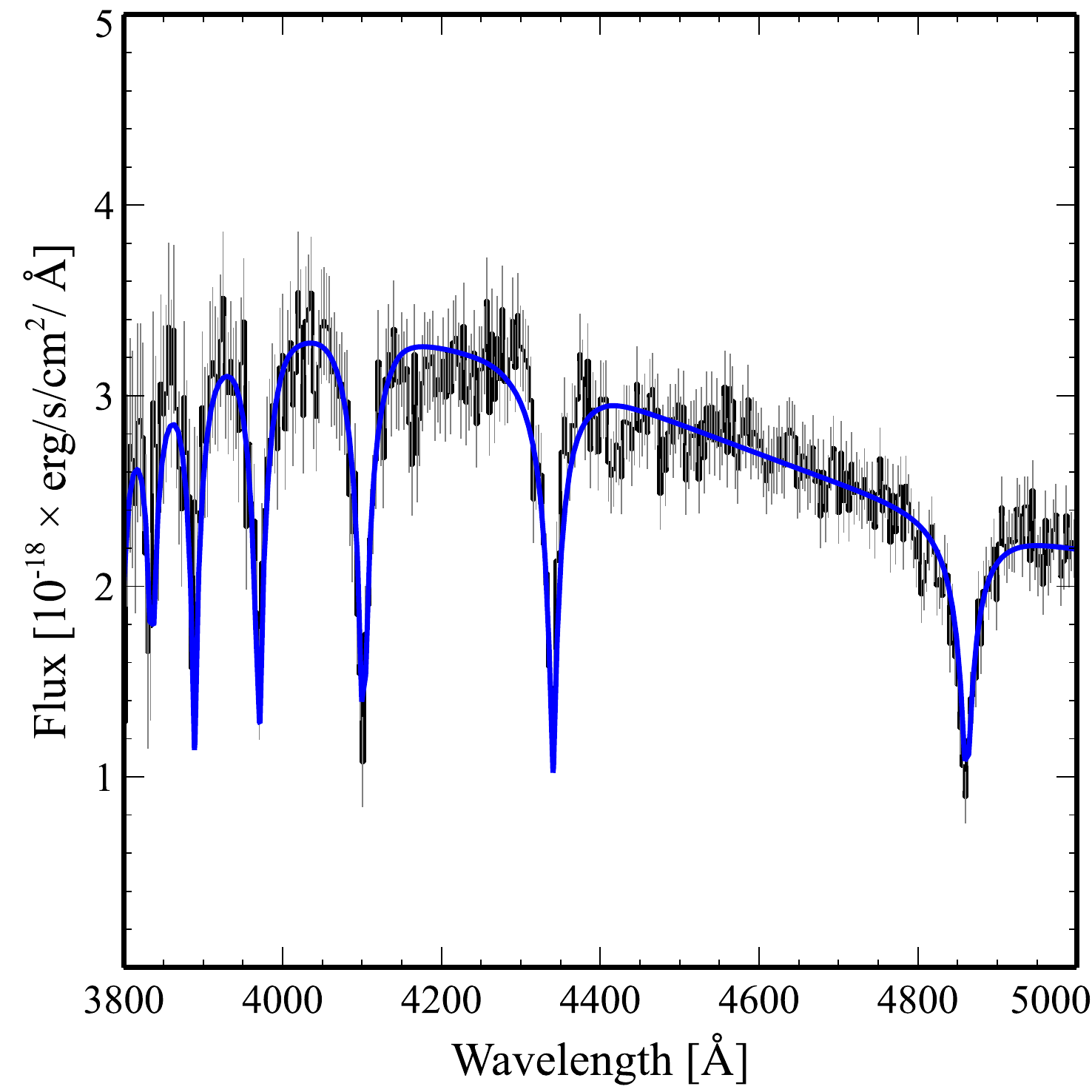}
\caption{Zero-velocity median combination of the spectra shown in Figure\,1, (binned by 4 for clarity). The solid blue line shows the best-fit spectrum. Measurement uncertainties are shown in grey.}
\end{center}
\end{figure}
 
The derived atmospheric parameters place the WD in a regime where convective atmospheric effects are known to produce systematic errors in 1D model atmospheres \citep{tls+11}. Using the numerical estimates of \cite{tls+13} and \cite{tgk+15}, we find that the ``true'' atmospheric parameters are $T_{\rm eff} = 8600\pm190$\,K and $\log g = 6.97\pm 0.22$\,dex.

\subsection{WD radius and model-independent surface gravity} 
Using the temperature estimate, the observed flux, and the parallactic distance, we can obtain an estimate of the WD radius via, 
\begin{multline}
m_{\lambda} - 5\log_{10}(d/10\,{\rm pc}) - A_{\lambda} = \\-5\log(R_{\odot}/10\,{\rm pc}) - 5\log_{10}(R/R_{\odot}) - 2.5\log_{10}F_{\lambda} + c_{\lambda}, 
\end{multline}
where $m_{\lambda}$ is the apparent magnitude in band $\lambda$, A$_{\lambda}$ is the corresponding reddening, F$_{\lambda}$ is the emitted \replaced{flux}{luminosity} per unit area integrated over the bandpass and $c_{\lambda}$ is the zero-point of the filter. 
Convolving the best-fit atmospheric model with the SDSS-$g$ bandpass \citep{sdss_filt}\footnote{http://svo2.cab.inta-csic.es/svo/theory/fps3/index.php?id=SLOAN/SDSS.g} yields F$_{\rm g} =4.9745\times10^{7}$\,erg\,cm$^{-2}$\,s$^{-1}$.
For the reddening, the galactic-extinction map of \cite{sfd98} gives $A_{\rm g} = 0.481$ for the total extinction along the line of sight, which however can be considerably smaller given the proximity of the system. If we conservatively adopt $A_{\rm g} = 0.16 - 0.481$ we find, 
 $R_{\rm WD} = 0.024^{+0.004}_{-0.002}$\,R$_{\odot}$, where the uncertainty also takes into account the parallactic and photometric errors. 

Together with the WD mass derived in Paper\,II, the former estimate yields $\log_{10} g = 7.11^{+0.08}_{-0.16}$\,dex, for the surface gravity, consistent with the previous spectroscopic estimate (Section\,2.2). 
\subsection{Kinematics}
The 3D velocity information from the combined timing and spectroscopic analysis allows to calculate the orbit of the system inside the gravitational potential of the Galaxy. Figure\,3 shows the system's  orbit based on the \cite{kbg+08} empirical model for the Galactic potential, over the past 1.5\,Gyrs. We find that the orbit of the system is highly eccentric with a galactocentric distance varying from 4 to 15\,kpc and a vertical component extending to $\sim 2$\,kpc. The  velocity relative to the local standard of rest  when the system crosses the galactic plane ($Z=0$) ranges from $\sim 55$ to 130\,km\,s$^{-1}$. In Section\,4 we discuss the implications of these constraints for the formation of the system. 

\begin{figure*}
\begin{center}
\includegraphics[width=1.0\textwidth]{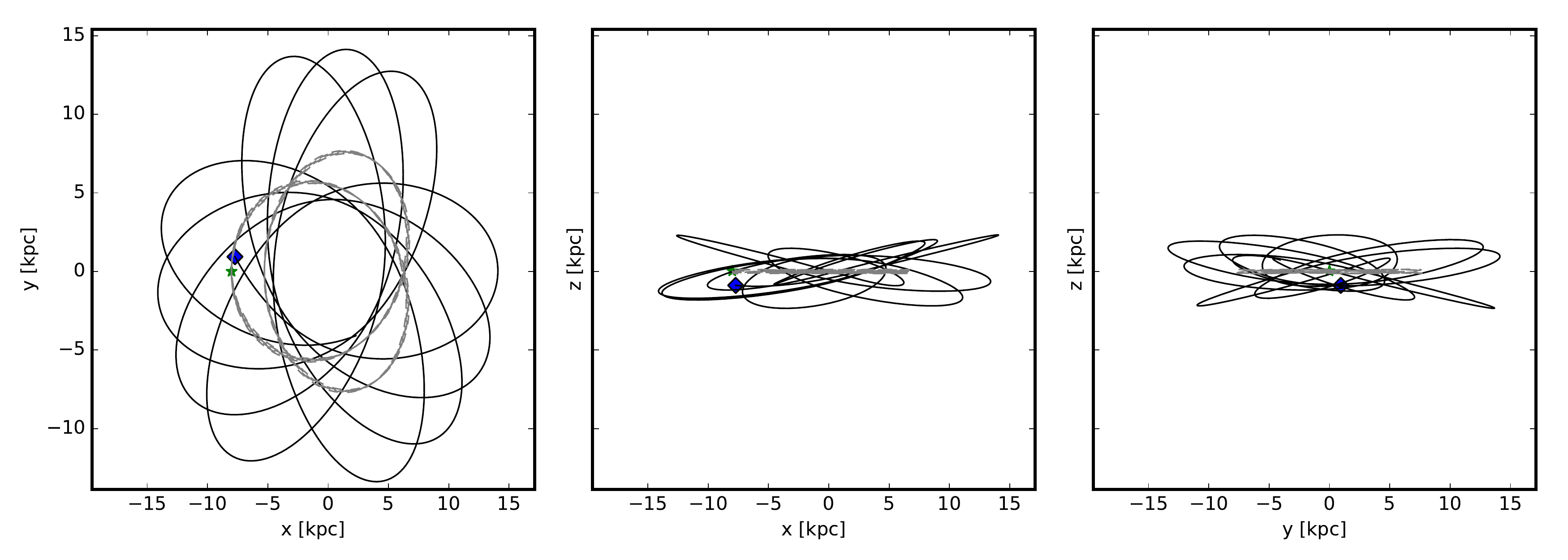}
\caption{3D motion of PSR\,J2234+0611 in the Galaxy over the past 1.5\,Gyr, based on the potential of  \cite{kbg+08} (solid line). The dashed line shows the orbit of the Sun. The current positions of \psr and the Sun are marked with a diamond and a star respectively. All axes are in kiloparsecs. }
\end{center}
\end{figure*}

\subsection{Spectrophotometry}
Finally, we computed synthetic magnitudes by convolving our spectra with the SDSS $g$-filter response. Uncertainties were estimated using a Monte-Carlo approach, where multiple  realizations of the spectra were convolved and compared with each other. 
\begin{figure}
\begin{center}
\includegraphics[width=0.45\textwidth]{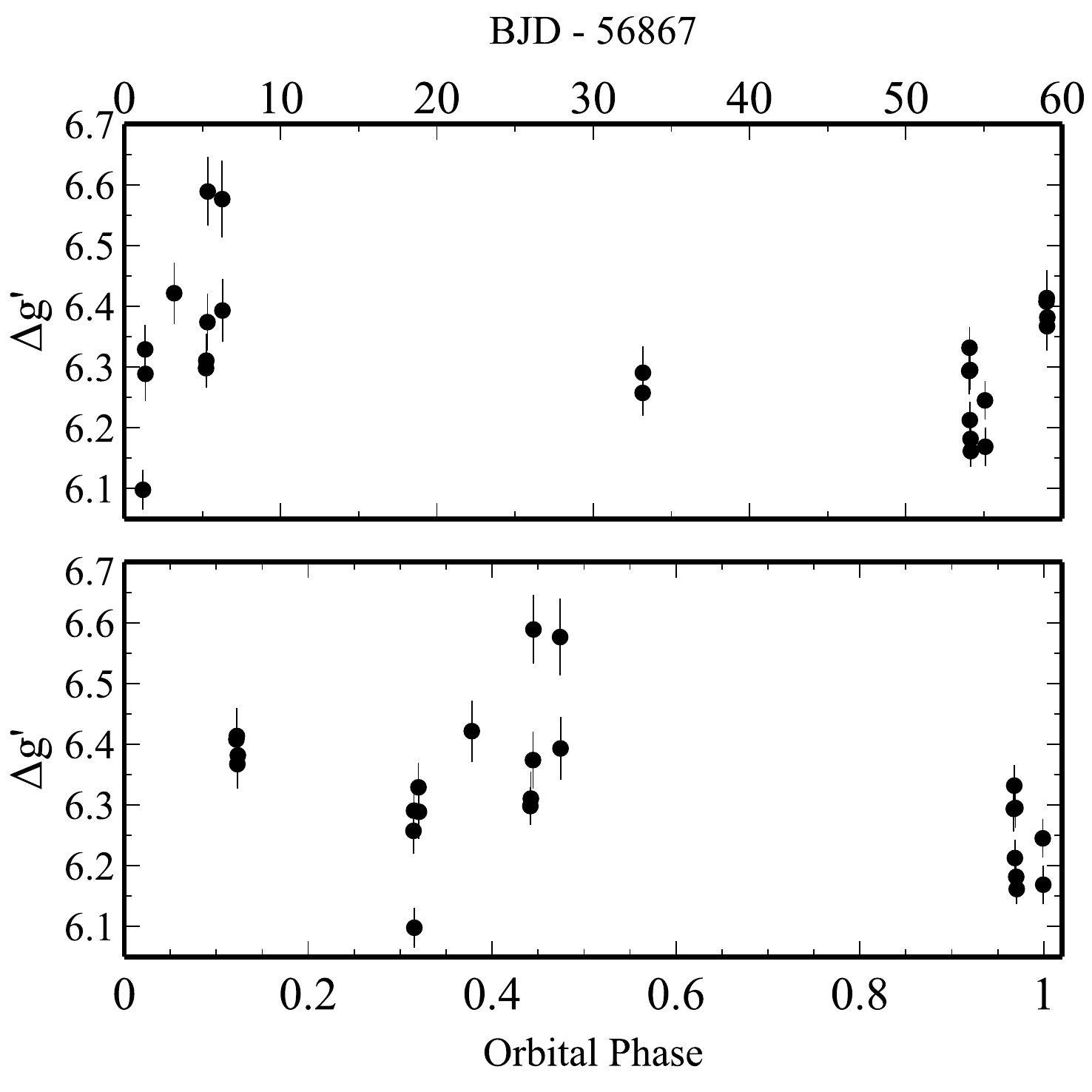}
\caption{g-band differential synthetic magnitudes for the companion to PSR\,J2234+0611 displayed as a function of time (upper) and orbital phase (lower)}
\end{center}
\end{figure}
The differential magnitude relative to the reference star is shown in Figure\,4.  The observed peak-to-peak difference of $\sim 0.5$\,mag, is more than an order of magnitude larger than the formal uncertainties. We find no compelling evidence for correlation with the orbital motion. Given the differential refraction effects which likely polluted our radial velocities, it is possible that part of the scatter can be accounted for by slit displacements. However,  we find no definite correlation with this effect. Another possible cause  is an intrinsic luminosity change of the LMWD which could be due to pulsational instabilities, like those seen in ZZ-Ceti stars and extremely low-mass WDs \citep[e.g.][and references therein]{khg+15}. 
\section{Discussion}
The observed and derived physical parameters of the system are summarized in Table\,2.
PSR J2234+0611 is the first known case of a Galactic eMSP with a LMWD companion. The combination of precision timing measurements (Paper\,II) and phase-resolved spectroscopy make the binary a unique test-bed for stellar  evolution physics. In the remainder of this Section we explore the ramifications of our work for LMWD models and place constraints on the evolution of the system.

\subsection{Origin and Evolution} 
One question we can address directly  is whether \psr evolved from an interacting binary. First, the WD nature of the companion is consistent with the expectations for LMXB evolution. For systems within  this orbital-period range,  the mass transfer starts while the donor star ascends the red-giant branch. For solar-metallicity progenitors, the theoretical mass-orbital period relation of \cite{ts99a}  predicts a mass of 0.29\,M$_{\odot}$ for the WD,  which is slightly larger than the observed value. However, the marginal difference  could be attributed to other factors such as initial ZAMS composition and/or remaining uncertainties in the evolution models of LMXBs. 

A second diagnostic is the system's motion in the Galaxy. PSR\,J2234+0611 has a peculiar velocity of $\sim 31$\,km\,s$^{-1}$, relative to the local standard of rest, which is typical for LMXB descendants \citep{lor05}. 
Furthermore, the ``crossing velocity'' (at Z=0) ranges from 34 to 130\,km\,s$^{-1}$, which is in broad agreement with the predictions for core-collapse supernovae \citep[e.g.][and references therein]{skj+06}. 

\added{Therefore, even though  a triple-origin scenario cannot be ruled out conclusively, the consistency of the properties of the PSR\,J2234+0611 system with Galactic-field MSPs strongly support the binary evolution hypothesis.}
\deleted{Overall, we believe that these properties strongly support the binary evolution hypothesis.} 

In regards to the specific mechanism that gave rise to the high eccentricity, further evidence come from the pulsar and WD masses. As we briefly discussed in Section\,1, one possibility is that the MSP formed via a spontaneous phase transition. Assuming a symmetric implosion with a negligible momentum kick,  the observed masses and eccentricity can be linked directly to the amount of mass radiated during the explosion, $\Delta M = e(M_{\rm PSR} + M_{\rm c}) = 0.215$\,M$_{\odot}$, and the progenitor mass, $M_{\rm PSR} + \Delta M =1.60$\,M$_{\odot}$. 
These constraints disfavour the scenario proposed by \cite{jldd15}, in which a NS transforms to a quark star, as we would expect such a transition to occur at high energy densities, relevant for NS masses $> 1.8$\,M$_{\odot}$ \citep[][c.f. Table\,1]{jldd15}. 

An alternative mechanism proposed by \cite{ft14} theorizes that eMSPs form  indirectly from a rotationally-delayed accretion-induced collapse (AIC) of a massive WD. First, 
because WDs require a fine-tuned mass transfer rate to grow in mass \citep{nk91,clxl11,tsyl13}, the resultant systems are expected to have orbital periods in the $10-60$\,days range, in agreement with the known eMSPs. \cite{ft14} propose that the AIC could in principle be delayed until after the cessation of mass transfer, due to the rapid rotation of the WD progenitor.  The radiated gravitational binding energy \added{during the AIC} would therefore induce the observed high eccentricities as  no further circularization is expected \added{from the two compact objects}.

To first order, the imploding WD should only be slightly above the Chandrasekhar limit \citep{nk91,dbo+06}, and therefore collapse into a pulsar with a gravitational mass in the 1.22-1.31\,$M_{\odot}$ range, assuming  small amounts of baryonic mass loss during the transition \citep{ft14}.  The constraint on the progenitor mass and the \replaced{relatively high systemic velocity derived here}{peculiar velocity of $\sim 31$\,km\,s$^{-1}$ inferred here}, therefore exclude this simple version of the RD-AIC mechanism. However, it is possible that if one relaxes the assumption for rigid rotation for the progenitor WD \citep{yl04}, the the mass of the resulting pulsar could be higher. In addition, \added{strong magnetic fields may also play an important role} \citep[see discussion in][]{ft14}.
\added{In such a case, an asymmetric mass loss would still be required to explain both the large systemic velocity and eccentricity \citep[][]{ft14}.}  

For the AIC to be successfully, the loss of angular momentum should  happen on a timescale larger than the Debye cooling time, to allow for efficient $^{24}$Mg and $^{20}$Ne electron captures instead of oxygen deflagration of the (hot) core \citep{tsyl13,ft14}. Since super-Chandrasekhar WDs crystalize within $\sim10^8$ \,yr \citep[e.g.][]{bsw95}, the inferred LMWD cooling age of $\sim 1.5$\,Gyr (see below) is therefore 
also consistent with this scenario. 
In summary, albeit fine-tuned, an AIC mechanism cannot be ruled out completely. Mass measurements for other eMSPs are necessary to further constrain this scenario.

\begin{table}
 \caption{Properties of the PSR\,J2234+0611 system}
\resizebox{8.5cm}{!}{
\begin{tabular}{lr}
\hline
Observed parameter & Value \\
\noalign{\smallskip}
\hline
\noalign{\smallskip}
Reference Epoch (MJD) &  56794\\ 
Time of ascending node (MJD) &  56794.09318642(9) \\
$\mu_{\alpha}$ (mas\,yr$^{-1})$&+ 25.3896(271)\\
$\mu_{\delta}$ (mas\,yr$^{-1})$&+9.4816(619)\\
Parallax, $\pi$ (mas)& 0.7423(279)\\ 
Orbital period, $P_{b}$ (days)& 32.001401609(14)\\ 
Eccentricity, $e$ & 0.1292740499\\
\noalign{\smallskip} 
\hline
Inferred parameter & Value \\
\noalign{\smallskip}
\hline
\noalign{\smallskip}
Pulsar mass, $M_{\rm{PSR}}$ (M$_{\odot}$) \dotfill & $1.39(1)$ \\[.8ex]
WD mass, $M_{\rm{c}}$ (M$_{\odot}$, spectroscopy) \dotfill  & $0.275(8)$\\
Temperature (K; 3D-corrected) \dotfill & 8600(190)\\ 
Surface gravity ($\log g$, 3D-corrected) \dotfill & 6.97(22)\\
Surface gravity ($\log g$, $\pi$ + photometry)\dotfill & $7.11^{+0.08}_{-0.16}$\\[.8ex]
Photometry, $g$-band \dotfill & 22.17(10) \\ 
Semi-amplitude of radial velocity, $K_{\rm{WD}}$ (km\,s$^{-1}$) \dotfill & 81(23)\\
Systemic radial velocity, $\gamma$ (km\,s$^{-1}$) \dotfill & $-$20(34)\\
Transverse velocity, $v_T$ (km\,s$^{-1}$) \dotfill & 179\\
3D velocity amplitude (km\,s$^{-1}$) \dotfill & 180  \\ 
Mass ratio, $q$ (timing) \dotfill & 5.05\\
WD radius (Photometry) (R$_{\odot}$) \dotfill & $0.024^{+0.004}_{-0.002}$ \\[.8ex]
Cooling age, $\tau_{\rm{c}}$ (Gyr) \dotfill & 1.5 \\[.8ex]
\noalign{\smallskip} 
\hline
\end{tabular}
}
\label{table:2}
\end{table}

Finally, we consider the possibility of eccentricity pumping via a short-term interaction between the post-LMXB system and a circumbinary (CB) disk  \citep{ant14}. Such a disk can be fuelled by material escaping the proto-WD due unstable CNO burning (H-flash). Because H-flashes are expected only for a limited range of WD masses \citep[$\sim 0.2 - 0.35$\,M$_{\odot}$, e.g.][]{amc13,antoniadis,itla14}, this mechanism predicts a statistical correlation between the eccentricity and orbital period which is  applicable to  \emph{all} MSP systems. This indeed seems to be the case, as all eMSPs, including PSR\,J2234+0611, have  orbital periods between 22 and 32 days  (c.f. Figure\,5 and references in the caption) --- a regime where circular MSPs have yet to be discovered. 
In the analytic framework considered by \cite{ant14}, the observed eccentricities are linked to the CB-disk mass and lifetime, as well as the initial eccentricity \citep{acl91,dijv13}. This work finds that an eccentricity of $e\sim 0.13$ for the observed orbital separation requires $M_{\rm d}\tau_{\rm d}\simeq 75$\,M$_{\odot}$\,yr for the disk mass and lifetime. For the typical mass-loss of $10^{-4}$\,M$_{\odot}$ occurring during an H-flash \cite{ant14}, this yields $\tau_{\rm d}\simeq 50\,000$\,yr, which is much shorter than the inferred cooling age. In addition, it is possible that the interaction is even more efficient, if the dependence on the eccentricity of the disk's central cavity is weak, as found in recent high-resolution shock capturing simulations \citep{dhm13}. 

\begin{figure}
\includegraphics[width=0.45\textwidth]{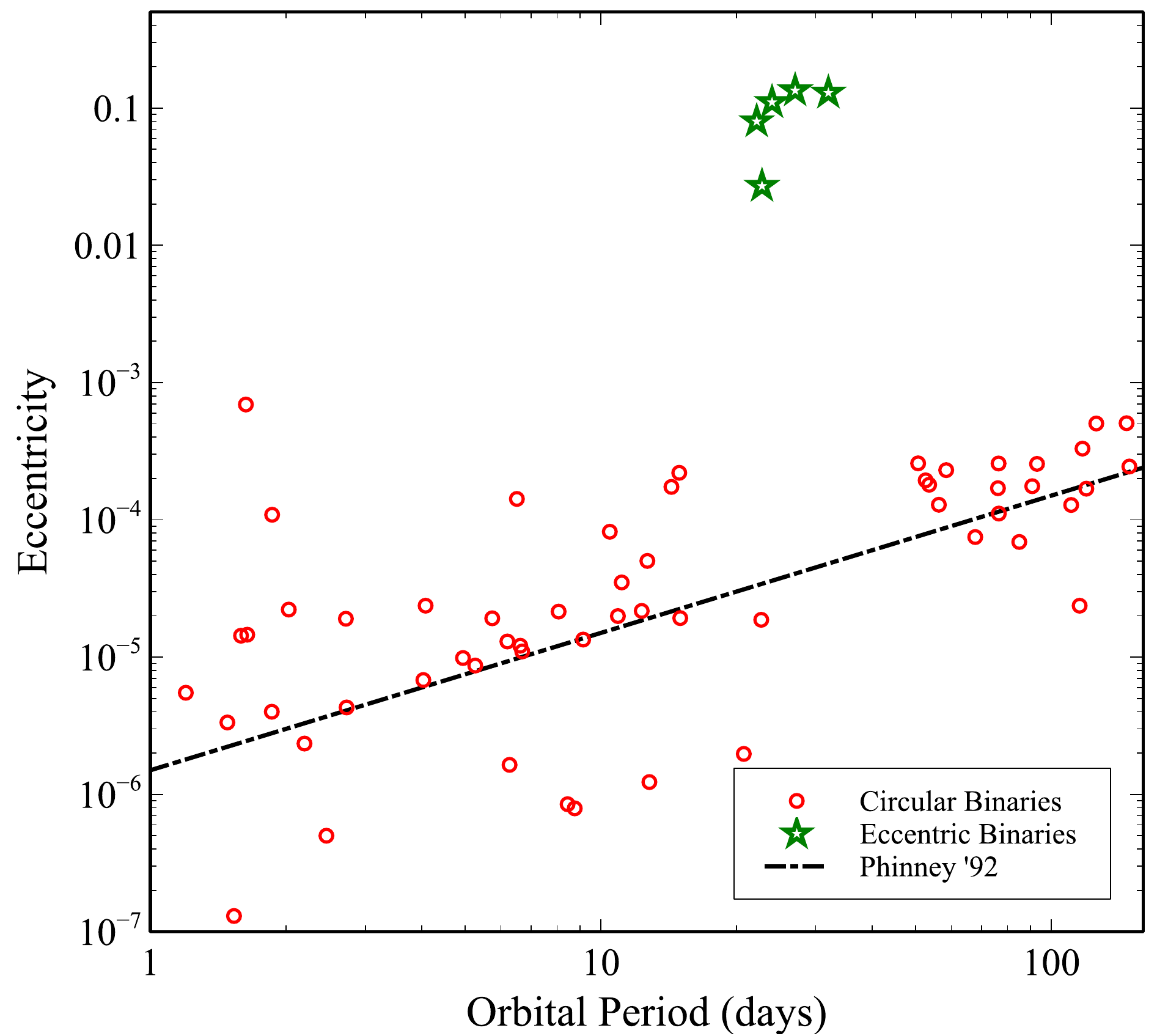}
\caption{Orbital periods and eccentricities of binary MSPs with orbital periods between 1 and 150\,days. Circular binaries (red circles) are taken from the ATNF pulsar catalogue \citep{mhth05}. The eMSPs shown are (from left to right) PSRs J1618$-$3919 \citep{bai10}, J0955$-$6150 \citep{ckr+15}, J1950+2414 \citep{kls+15}, J1946+3417 \citep{bck+13} and J2234+0611 \citep[][this work]{dsm+13}. The theoretical prediction of \cite{phi92} is also shown as a dashed line. Note that the apparent increased scatter at small orbital periods may be artificially induced by covariances between the eccentricity and Shapiro-delay in the timing model.}
\end{figure}
The CB-disk mechanism poses no additional constraint on the eMSP masses, apart from those expected from nuclear, core-collapse and accretion physics \citep{sew+15}. 

The observed pulsar mass of 1.39\,M$_{\odot}$ implies a small accretion efficiency for the  recycling process. Following \cite{avk+12}, if we indeed interpret the system as a direct descendant of a sub-Eddington mass-transferring binary, the donor's initial mass must have been $\gtrsim 1$\,M$_{\odot}$, in order for the star to reach the red-giant branch within a Hubble time. Assuming a pulsar birth mass $\geqslant 1.17$\,M$_{\odot}$ \citep{msf+15}, the former considerations therefore imply an efficiency of $\alpha < 30\%$. A more ``typical'' NS birth mass of $\sim 1.35$\,M$_{\odot}$ \citep{opn+12} however, yields $\alpha < 6\%$.

\subsection{Prospects for LMWD physics}
PSR\,J2234+0611 is only the third pulsar-LMWD binary with model-independent  mass and radius constraints \citep[see ][for  earlier discussions]{avk+12,antoniadis,kvk+14}, making the system a valuable test-bed for WD atmospheric and cooling properties.   Unfortunately, the  precision of our spectroscopic measurements does not suffice for a detailed confrontation with theoretical model atmospheres. 
For instance, both the 1D  and 3D-corrected atmospheric properties are consistent with the model-independent estimates within formal uncertainties (see Sections 3.2 \& 3.3).  Further spectroscopic observations, which will increase the signal-to-noise of the spectrum, are therefore required to draw any further conclusions. 

The timing mass and radius constraints are consistent with cooling models for WDs with relatively thin hydrogen atmospheres. For instance, the recent models of \cite{amc13} yield a radius of $R = 0.022$\,R$_{\odot}$ for a 0.275\,M$_{\odot}$ WD, which is in perfect agreement with our estimate derived in Section\,3.3. The same models yield a cooling age of $\tau_{\rm c} = 1.5$\,Gyr. The former should also be comparable to the age of the system as the expected evolution timescale of a 0.275\,M$_{\odot}$ proto-WD before it settles on the cooling branch is small \citep{itla14}.

Perhaps more important is the fact that PSR\,J2234+0611 lies close to the low-mass extension of the ZZ-Ceti instability strip, as derived empirically by \cite{gkb+15}. 
 The  variability of $\sim0.5$\,mag seen in our dataset is large compared to what is found for other pulsating LMWDs, like for instance PSR\,J1738+0333 \citep{khg+15}, and more similar to classical ZZ-Ceti stars. 
 If PSR\,J2234+0611 indeed pulsates due to excitation of non-radial $g-$modes \citep{crah12,vfbd13}, we would expect a longer dominant periodicity due to the lower temperature compared to PSR\,J1738+0333 \citep{avk+12}.   
 Further high-cadence photometric observations have the potential to probe the WD interior in detail and help infer (and calibrate) the atmospheric composition, hydrogen mass and interior convective properties. Even if pulsations are not confirmed the system will place strong constraints on the exact location of the instability strip in a poorly constrained regime.

\section{Conclusions} 
We have presented phase-resolved spectroscopic observations of the companion to PSR\,J2234+0611. Our data unabiguously identify the star as a low-mass He WD ---the first found orbiting a galactic-field MSP with non-zero eccentricity. 

We find that the WD mass is consistent with the expectations for LMXB evolution and strongly disfavours a triple-star formation hypothesis. 
In addition, the pulsar mass (Table\,2) contradicts  the quark-nova formation theory proposed by \cite{jldd15}. Combined with the \replaced{high-systemic velocity}{inferred peculiar velocity}, it also poses stringent constraints on the rotationally-delayed AIC hypothesis of \cite{ft14}, as the latter requires both a differentially rotating super-Chandrasekhar mass WD progenitor and asymmetric mass-loss at birth. 
On the other hand, we find the mechanism of \cite{ant14} --which proposes  eccentricity pumping via interaction with a transient CB disk-- to be consistent with all observed and inferred parameters. If PSR\,J2234+0611 indeed originates from a LMXB, the low pulsar mass implies a small accretion efficiency during recycling of at most 30\%, with a more likely value close to $6\%$. 

Finally, we find tentative evidence for pulsations, which together with the independent constraints on the stellar radius and mass, transform the system into a unique test-bed for LMWD evolution\deleted{and mixing-length theory}. We are looking forward for further detailed spectroscopy and high-cadence photometry that will allow for a detailed and unprecedented confrontation with models. 

\acknowledgements 
This work is based on observations made with ESO Telescopes at the Paranal Observatories
under programme ID\,093.D-0108(A). We wish to thank the ESO astronomers-on-duty for the excellent execution of the observations. 

JA is a Dunlap Fellow at the Dunlap Institute for Astronomy and Astrophysics at the University of Toronto.  The Dunlap Institute is funded by an endowment established by the David Dunlap family and the University of Toronto.  P. C. C. F. PF acknowledges
financial support by the European Research
Council for the ERC Starting grant AST 1108753. 
J. S. D. was supported by the NASA's Fermi Guest Investigator program. 

We have made extensive  use of NASA's Astrophysics Data System and Astropy, a community-developed core Python package for Astronomy.

\bibliographystyle{apj}
\bibliography{author}

\end{document}